\documentclass{article} 
\def \inbar{\vrule height1.5ex width.4pt depth0pt} 
\def \C{\relax\hbox{\kern.25em$\inbar\kern-.3em{\rm C}$}} 
\def \R{\relax{\rm I\kern-.18em R}}

\newcommand{\sgn}{{\rm sgn}}

 \newcommand{\beq}{\begin{equation}} 
\newcommand{\eeq}{\end{equation}}
 \newcommand{\bea}{\begin{eqnarray}}
 \newcommand{\eea}{\end{eqnarray}} 
\newcommand{\nn}{\nonumber}

\newcommand{\Tr}{\hbox{Tr}} 
\newcommand{\Str}{\hbox{Str}}

\begin{document} 
\author{Erdal Toprak${\,}^{2}$ 
and  O. Teoman Turgut${\,}^{1,2,3}$\\  \\ ${}^{1}\,
$Department of Physics, KTH\\
SE-106 91 Stockholm, Sweden\\ and 
\\${}^{2}\,
$Department of Physics, Bogazici University \\ 
80815 Bebek, Istanbul, Turkey\\  
and\\ ${ }^{3}$Feza Gursey Institute\\ 
Kandilli 81220, Istanbul, Turkey\\   turgutte@boun.edu.tr\\
turgut@theophys.kth.se}   
\title{\bf  Large N Limit of SO(N)  Gauge Theory of Fermions and Bosons } 
\maketitle 
\large 
\begin{abstract} 
\large  
In this paper we study the large $N_c$ limit of $SO(N_c)$ gauge theory
 coupled to 
a Majorana field  and a real scalar field 
in $1+1$ dimensions extending ideas of Rajeev\cite{2dqhd}.
We show that the phase space of the  resulting classical theory 
of bilinears, which are the mesonic operators  of  this theory, is  
$OSp_1({\cal H}|{\cal H} )/U({\cal H}_+|{\cal H}_+)$, 
where ${\cal H}|{\cal H}$ refers to the underlying 
complex  graded space of combined one-particle states of 
fermions and bosons  and 
${\cal H}_+|{\cal H}_+$  corresponds to the positive frequency 
subspace. In the begining to simplify our  presentation we 
discuss in detail the case with Majorana fermions only
(the purely bosonic case is treated in \cite{erdal}).
In the Majorana fermion  case the phase space
is given by $O_1({\cal H})/U({\cal H}_+)$,
where ${\cal H}$ refers to the complex one-particle
states and ${\cal H}_+$ to its positive frequency subspace.
The meson spectrum in the linear approximation 
again obeys a variant of the 't Hooft equation.
The linear 
approximation to the boson/fermion coupled case brings 
an additonal bound state equation for mesons, 
which consists of one fermion and one boson, 
again of the same form as the well-known 
't Hooft equation. 

\end{abstract} 
\large 
\section{Introduction}

Gauge theories play a fundamental role in our description  of nature.
Nevertheless our understanding of confining phase of gauge theories is 
not so complete. 
In principle we should be able to calculate  the hadronic spectrum
starting from Quantum Chromodynamics(QCD), which is a gauge theory,
yet this has not been possible up to now.
It is believed that the hadrons are colorless 
excitations of the underlying gauge theory and we never see 
the constituent quarks as free particles.  This suggests that in this 
case we should have an independent formulation of gauge theories 
in terms of color singlet operators of the original gauge theory.
In general this is a very hard task.

Gauge theories in $1+1$ dimensions   provide a great 
testing ground for many ideas about realistic theories.
This  is a great simplification,
various difficult problems of higher dimensional theories will not be there,
yet there are still interesting aspects of these theories which make 
them worth studying in depth. 
In \cite{2dqhd} Rajeev has constructed a theory of mesons in two dimensions in 
the limit $N_c$, the number of colors in $SU(N_c)$, goes  to infinity
using only the color invariant variables (which correspond to the meson
operators). The idea that  
QCD should  simplify while keeping all its essential features  in this limit  
goes back to `t Hooft\cite{thooft1, thooft2} and that this limit  
should be a kind of classical  
mechanics to Migdal and Witten \cite{yaffe}. 
This is a very promising  step in simplifying gauge theories,
but  the large-$N_c$ theory is also quite 
complicated and it is not possible as yet to understand it 
in four dimensions. 

Originally `t Hooft studied  two dimensional 
QCD  in the large-$N_c$ limit  to understand the meson spectrum 
and obtained his bound state  
equation in his seminal  paper\cite{thooft2}. 
Soon after the scalar two dimensional QCD was worked out by Shei and Tsao  
in \cite{shei} following `t Hooft, and later by Tomaras using Hamiltonian 
methods in \cite{tomaras}. 
These works obtained the analog of the `t Hooft equation for this case. 
A natural extension of these would be to look at 
combined (fermionic) QCD and scalar QCD, 
 this is done  
in a paper of Aoki\cite{aoki1} where it is shown that three  types of  
mesons are possible and they all obey   
a certain type of `t Hooft equation(see also \cite{aoki2}).    
Cavicchi \cite{cavicchi} 
using a path integral approach with   
bilocal fields, developed in \cite{divecchi},
studied  coupled fermions and bosons as well as some 
other  models in two dimensions and he obtained 
some generalized versions of the 't Hooft equation. 

To understand 
 gauge theories better,   we study the  problem of
bosons and fermions coupled to $SO(N_c)$ gauge fields in $1+1$
dimensions.  
 We will apply the methods  developed  by Rajeev 
to this toy model. We recommend his  lectures  for a more detailed  
exposition of the underlying ideas  and various other 
directions \cite{istlect}. 
In  \cite{2dqhd} it was shown that the phase space of the two dimensional QCD is an  
infinite dimensional Grassmannian\cite{pressegal}. 
Using  the same methods scalar version of QCD is worked out in  
\cite{rajtur},the  
phase space of the theory comes out to be  an infinite dimensional disc.  
Recently Konechny and the second author 
obtained the large-$N_c$ phase space of bosons and 
fermions coupled to $SU(N_c)$ gauge theory;  
a certain kind of super-Grassmannian \cite{tolyateo}. 
The linearized equations agree with the ones found in \cite{aoki1}. 
The correct  
equations are nonlinear and various approximation schemes are 
also discussed in \cite{tolyateo}.
There are some  
ideas in the literature which suggest that gauge theories in 
two dimensions all behave in a  
very similar way\cite{kutasov}, therefore it will be 
interesting to see how much of this holds for 
$SO(N_c)$ gauge theory. 

The organization of our work is as follows, since we did not 
want to go into technical details of super-geometry immediately,
we first study the purely fermionic case.
The essential calculations  are very similar to the ones 
in Rajeev's lectures\cite{istlect} 
and for the geometry basic ideas are 
already in \cite{bowraj,  pressegal}, we also 
recommend the article \cite{gracia} for 
a good discussion. We show that one can 
formulate the large-$N_c$ limit in terms of 
bilinears along the lines in \cite{2dqhd}.
We obtain a variant  of the
't Hooft equation in the linear approximation.
We explain the geometry of the phase space and show that 
it is a homogeneous manifold, $O_1({\cal H})/U({\cal H}_+)$(see 
explanations in section IV), and the symplectic form is the 
natural one.
In the second part, we study the combined system of bosons and fermions,
this part is very brief, we state mostly the results. 
We obtain a super-Poisson structure of the 
bilinears in the large-$N_c$ limit and the resulting Hamiltonian. 
The equations of motion in the linear approximation agree with the
purely bosonic and purely fermionic ones with an additional one for 
the mesons made up  of 
one fermion and one  boson. This is again a variant of the well-known 
't Hooft equation. 
The discussion on the geometry of the resulting  
infinite dimensional supersymplectic space requires 
some new ideas. This part is technically complicated, we use 
essentially Berezin's ideas \cite{berezin}, but we do not claim 
that all the technicalities of the infinite dimensional case is 
understood. We show that the 
underlying phase space should be the super-homogeneous manifold 
$OSp_1({\cal H}|{\cal H})/U({\cal H }_+|{\cal H}_+)$, and the 
supersymplectic form is the natural one on this space.
We plan to come to the more mathematical aspects of this problem in a 
future publication.

\section{The $SO(N_c)$ Majorana Fermions in the Light-cone}
  
Since the basic philosophy was explained in \cite{2dqhd} we can be 
brief and only  state our conventions and 
define our theory.  
 We will use the light cone coordinates  
$x^{+} = \frac{1}{\sqrt{2}}( t+x)$,  $x^{-} = \frac{1}{\sqrt{2}}( t-x)$,
(we recommend 
\cite{heinzl} for an introduction to light-cone quantization, 
and \cite{brodsky} for a more comprehensive review),
the action functional is    
\begin{eqnarray} 
S =  \int \bigl[{1 \over 2}\Tr F_{\mu\nu}F^{\mu\nu} +i\bar
\Psi_M \gamma^\mu(\partial_\mu+gA_\mu)\Psi_M- m\bar \Psi_M
\Psi_M\bigr] 
,\end{eqnarray} 
where we have an $SO(N_c)$ gauge theory for which  the matter fields 
are in the fundamental representation and Tr denotes an invariant inner product in the 
Lie algebra.  
The Lie algebra condition for $SO(N_c)$ implies that $A_\mu^T=-A_\mu$.
To compute the variations of  the action 
we need the independent degress of freedom,
we can expand $A_\mu=A_\mu^aT^a$ where $T^a$ are 
the generators of the Lie algebra
of $SO(N_c)$, chosen such that  $\Tr T^a T^b=-{1\over 2}\delta^{ab}$.
Our  conventions for the Majorana fermions are as follows:
we choose the Majorana representation in which the fermions are real, i. e. 
$\Psi_M^\dag=\Psi_M^T$(transpose here also includes   the color indices
to simplify the notation). 
The gamma matrices now are given by,
\bea
   \gamma^0=\pmatrix{0&-i\cr i&0}\quad \gamma^1=\pmatrix{0&-i\cr -i&0}
\quad \gamma^5=\pmatrix{-1&0\cr 0&1}
.\eea
Note that $\gamma^5$ happens to be diagonal in $1+1$ dimensions
 and we set   $\bar \Psi_M=\Psi_M^T\gamma^0$.
We now rewrite the action  in the light-cone coordinates
and eliminate all nondynamical degrees of freedom.
We write  $\Psi_M=\pmatrix{\psi_1\cr \psi_2}$, and 
use  $\gamma^+={1\over \sqrt{2}}(\gamma^0+\gamma^1)$, 
$\gamma^-={1\over \sqrt{2}}(\gamma^0-\gamma^1)$.
We further set $A_-=0$ and choose $x^+$ as the evolution variable, which 
we call  ``time".
\beq
 S=\int dx^+dx^-\bigl[{1\over 2} (\partial_- A_+^a)^2+i\sqrt{2}\psi_1^T
\partial_-\psi_1
+i\sqrt{2}\psi_2^T\partial_+\psi_2+i2m\psi_1^T\psi_2+
i\sqrt{2}g\psi_2^TA_+^aT^a\psi_2\bigr]
,\eeq 
(from now on $T$ only means  tranpose in the color space).
We note also that we have a {\it real } two component fermion, they are 
Grassmann valued obeying  $\psi_1\psi_2=-\psi_2\psi_1$. We can check 
that the action is real if we use the following complex conjugation 
convention for spinors, $(\psi\xi)^*=\xi^*\psi^*$.   
We see that $\psi_1^\alpha$ is non-dynamical,
and hence can be eliminated using its equation of motion,
\beq 
    \psi_1^\alpha=-{m\over \sqrt{2}\partial_-}\psi_2^\alpha
.\eeq
Similarly we solve for the nondynamical $A_+^a$, and get
\beq
   A_+^a={i\sqrt{2} g\over \partial_-^2}\psi_2^TT^a\psi_2
.\eeq
A remark is in order here to clarify 
what we mean by `real fermions' while 
the action and the constraint equation for $A_+^a$ have
 explicit factors of $i$.   The resolution of this 
seeming paradox    is that it is the equations of motion 
which are actually real. To see this note 
first that the symplectic form has a factor of $i$ in it, and 
the 
$\psi_1^\alpha$ constraint  only has real 
operators. The $A_+^a$ constraint is also real if we choose 
complex conjugation of fermions to be
$(\psi\xi)^*=\xi^*\psi^*$. This convention implies  that the 
product of two real fermions is imaginary, and this  is the reason for the 
 extra factors of $i$ in the action.
The equation of motion for $\psi_2^\alpha$ reads,
\beq 
  \partial_+\psi_2^\alpha={m\over \sqrt{2}}\psi_1^\alpha
,\eeq
which is manifestly real.
This shows that the ``time" evolution  preserves the 
real valuedness  condition imposed on the 
fermions.

If we insert  the above constraints   into the action we arrive at
\beq
S =  \int dx^{+}dx^{-}\bigl[ i\sqrt{2}\psi_2 \partial_+\psi_2-
{\sqrt{2}m^2\over 2}
\psi_2{1\over i\partial_-}\psi_2-{g^2} \psi_2^T
T^a\psi_2{1\over \partial_-^2}\psi_2^TT^a\psi_2 \bigr]  
.\eeq 
This defines our theory at  the classical level with  
the redundant degrees of freedom eliminated.
{\it Since it is written entirely in terms of $\psi_2$ we will refer to 
this field as $\psi$ from now on}.
The real fermions have a super-poisson 
bracket, which can be read off from the 
action,  given by 
\beq
   \{ \psi^\alpha(x^-, x^+), \psi^\beta(y^-,x^+)\}_+={-i\over 2\sqrt{2}}
\delta^{\alpha\beta}\delta(x^--y^-)
.\eeq
Since our real fermions are Grassmann valued we use 
a symplectic structure which is $i$ times a real symmetric
operator, and the Hamiltonian 
is actually $i$ times an antisymmetric one,
as we will see in more detail in the next section.
There is an ambiguity in the 
quantization, we follow Rajeev's original 
approach\cite{istlect,2dqhd}, we will  remove the nondynamical fields 
 after quantizing the dynamical field,
$\psi^\alpha$.
Using the Dirac rule we get an anticommutator for $\psi$,
\beq
    [\psi^\alpha(x^-), \psi^\beta(y^-)]_+=i\hbar({-i\over 2\sqrt{2}}
\delta^{\alpha\beta}\delta(x^--y^-))=
{\hbar \over 2\sqrt{2}}\delta^{\alpha\beta}\delta(x^--y^-). 
\eeq
(Note that for the orthogonal group, the distinction  of upper and
lower indices is irrelevant, since the metric tensor is 
unity). The reason for our convention of complex conjugation  is 
to arrive at this more familiar form of the Clifford algebra. We could have 
chosen a convention in which the product of real fermions is
real, then we would  arrive at a Clifford algebra with a factor of $i$
as in \cite{bowraj}, but  this usual form
is preferable (it leads to a positive inner product, 
or the hermitian conjugation
is compatible with the inner product in the fermionic Fock space).
Let us introduce the Fourier decomposition, which is 
done in a complex Hilbert space (to simplify notation
we  drop the subscript in 
$p_-$, set $\hbar=1$  and   $[dp]={dp \over 2\pi}$), 
\beq
       \psi^\alpha(x^-)=\int_{-\infty}^{\infty}{[dp]\over 2^{3/4}}
\chi^\alpha(p)e^{-ipx^-}
.\eeq
(To be precise, in the above expansion, we should assume 
a cut-off $\epsilon_0$ around zero momentum, to be taken to zero at the end of
our calculations).
We see  that $\chi^\alpha(p)$ satisfies the basic 
anticommutator,
\beq
   [\chi^\alpha(p), \chi^\beta(q)]_+=\delta^{\alpha\beta}\delta[p+q]
,\eeq
where we write  $\delta[p-q]=2\pi\delta(p-q)$.
Real valuedness of the original field 
 implies that $\chi^{\alpha\dag} (p)=\chi^\alpha(-p)$.
In standard  physics notation the expansion 
would be written as
\beq
   \psi^\alpha(x^-)=\int_0^\infty[\chi^\alpha(p)e^{-ipx^-}+
\chi^{\alpha\dag}(p)e^{ipx^-}]
{[dp]\over 2^{3/4}}
.\eeq
As well known in the physics literature, to make the Hamiltonian 
bounded from below, we should choose a 
vacuum to be used to construct  a fermion Fock 
space, and further  impose a normal ordering prescription.
This is done by simply requiring  that 
$\chi^\alpha(p)|0>=0$ for $p > 0$, and defining
\beq
    :\chi^\alpha(p)\chi^\beta(q):=\cases{-\chi^\beta(q)\chi^\alpha(p)
\quad {\rm if}\quad p > 0,q<0,\cr
                            \chi^\alpha(p)\chi^\beta(q) \quad {\rm otherwise}}
.\eeq
This can be stated in one formula as,
\beq
    :\chi^\alpha(p)\chi^\beta(q):=\chi^\alpha(p)\chi^\beta(q)
-{1\over 2}\delta^{\alpha\beta}\delta[p+q][1+\sgn(p)]
.\eeq
For most of our calculations we need only 
the bilinears and the above expressions.
(For the Hamiltonian 
 we actually need the normal ordering of  product of four such operators,
and it is defined as usual all the annihilation operators are
to be taken  to the right  of creation operators, it will
be briefly explained later on).
We can reduce our Hamiltonian after this quantization process, and 
we get,
\beq
     H=\int dx^-\Big({m^2\sqrt{2}\over 2}:\psi^\alpha{1\over i\partial_-}
\psi^\alpha:-{g^2\over 2}:\psi^\alpha(x^-)\psi^\beta(x^-):|x^--y^-|
:\psi^\beta(y^-)\psi^\alpha(y^-):\Big)
.\eeq
Above we used 
$(T^a)^{\alpha\beta}( T^a)^{\lambda\sigma}=-{1\over 2}
(\delta^{\alpha\sigma}\delta^{\beta\lambda}-\delta^{\alpha\lambda}
\delta^{\beta\sigma})$ and the Green function
${1\over 2}|x^--y^-|=\partial_-^{-2}\delta(x^--y^-)$.
The last 
normal orderings can be 
rearranged  to act only on the color invariant 
combinations  in the large-$N_c$ limit, we will 
discuss  this in the next section.

Next we introduce the algebra of color invariant bilinears
and study the resulting system in the large-$N_c$ limit following 
\cite{2dqhd,istlect}.

\section{Classical mechanics of color invariant operators}

We define color invariant bilinears 
as in \cite{2dqhd, istlect, tolyateo} to be  our dynamical variables and 
find the large-$N_c$ limit by postulating 
Poisson algebra of these bilinears and 
defining the phase space to be a manifold where these 
Poisson brackets make sense. Since the theory is super-renormalizable
we expect this to be related to the Hilbert-Schmidt ideal condition 
which is well-known in the literature on the 
Fock spaces\cite{pressegal, mickbook, araki, plymen}.  We  will 
see  these aspects in more detail in 
 the next section when we talk about the geometry of the phase space.
We define our basic dynamical variables,  
bilinears,
\beq
     \hat R(p,q)={2\over N_c} \sum_\alpha :\chi^\alpha(p)\chi^\alpha(q):
,\eeq
which are color invariant combinations of the fermion operators.
We  find it useful to define a related operator,
$\hat F(p,q)=\hat R(-p,q)$, we will see that 
this is the correct variable for the 
geometry of the phase space.
We assume  that there are proper large-$N_c$ limits of our 
operators, then  they become classical variables when they are 
restricted to color invariant sector of the full Fock 
space. Following \cite{2dqhd}, we postulate the following 
Poisson brackets(we choose the quantization parameter to be
${1\over N_c}$),
\bea
\{ R(p,q), R(s,t) \}&=&-2i\Big(R(p,t)\delta[q+s]-R(q,t)\delta[p+s]
+R(s,p)\delta[q+t]-R(s,q)\delta[p+t]\nn\cr
&+&(\delta[q+t]\delta[p+s]-\delta[p+t]\delta[q+s])({\rm sgn}(t)+
{\rm sgn}(s))\Big).  
\eea    

Our dynamical system is not defined completely yet, since 
 there is still  a left over global color invariance,
generated by 
\beq
    \hat Q^{\alpha\beta}=\int [dp]:\chi^{\alpha\dag}(p)\chi^\beta(p):=
\int_0^\infty [dp] \chi^{\alpha\dag}(p)\chi^\beta(p)-
\int_0^\infty [dp] \chi^{\beta \dag}(p)\chi^\alpha(p)
.\eeq
The commutators of these 
generators satisfy the Lie algebra of $SO(N_c)$.

If we restrict ourselves to the color invariant states,
we find   a constraint equation satisfied 
in the large-$N_c$ limit, which can be best expressed in terms 
of $F(p,q)=R(-p,q)$,
\beq
     \int [dq] F(p,q)F(q,s)-{\rm sgn}(p)F(p,s)-
F(p,s){\rm sgn}(s)=0
.\eeq
We  define 
$\epsilon(p,q)=-{\rm sgn}(p)\delta[p-q]$,
then we can rewrite this constraint  as a simple quadratic operator equation,
\beq
     (F+\epsilon)^2=1
,\eeq
(we interpret $F,\epsilon$ as integral kernels acting 
on $L_2$ space of initial data).
In the next section we will analyze the geometric meaning of 
these constraints.
The  Hamiltonian and  the above Poisson brackets  
determine the 
evolution of our classical system; the Poisson brackets are
consistent with the constraint equation.

The large-$N_c$ Hamiltonian 
is obtained by dividing the original Hamiltonian by 
$N_c$ and rewriting it in terms of our large-$N_c$ variables.
After certain manipulations which are sketched below,
we obtain the following   Hamiltonian,
\beq
   H={\cal P}\int {1\over 8}(m^2-{g^2\over 2\pi}){[dp]\over p}R(-p,p)
-{g^2\over 64}{\cal FP}\int {[dpdqdsdt]\over (p+s)^2} R(p,q)R(s,t)
\delta[p+q+s+t]
,\eeq
where ${\cal P}$ and ${\cal FP}$ refer to the principal value and 
finite part prescriptions, respectively. In the following we will often 
write $\int$ short for ${\cal P}\int$ and ${\cal FP}\int$,
but one should keep in mind that these regularization perscriptions are 
used to define the singular integrals.
The main steps of the derivation of the above Hamiltonian
are very similar to the one in \cite{istlect}, 
although there are some small differences.
Here we supply the basic  ingredients 
to help the reader:
for simplicity in many places we write $x,y$ instead of $x^-,y^-$, 
we define $\epsilon(z)={\cal P}\int_{-\infty}^\infty {\rm sgn}(p)e^{+ipz}$,
note the sign of the exponent.
We have the vacuum expectation value of our field product,
\beq
<0|\psi^\alpha(x^-)\psi^\beta(y^-)|0>={1\over 4\sqrt{2}}
[\delta(x^--y^-)-\epsilon(x^--y^-)]
.\eeq
An important formula for 
the reduction is given in \cite{istlect}:
If $f(x,y)=\int[dpdq] e^{ipx+iqy}\tilde f(p,q)$,
\beq
   \int \epsilon(x-y)|x-y|f(x,y)=-{1\over \pi}{\cal P}\int {[dp]\over p}
\tilde f(-p,p) 
.\eeq
We also have $|x-y|={\cal FP}\int {[dp]\over p^2}e^{ip(x-y)}$.
We use  a form of Wick's theorem for normal  ordered products,
\bea
:\psi^\alpha(x)\psi^\beta(x)\!\!\!&::&\!\!\!\psi^\beta(y)\psi^\beta(y):=
:\psi^\alpha(x)\psi^\beta(x)\psi^\beta(y)\psi^\beta(y):
+<0|\psi^\alpha(x)\psi^\alpha(y)|0>:\psi^\beta(x)\psi^\beta(y):\nn\cr
&+&<0|\psi^\beta(x)\psi^\beta(y)|0>:\psi^\alpha(x)\psi^\alpha(y):
-<0|\psi^\alpha(x)\psi^\beta(y)|0>:\psi^\beta(x)\psi^\alpha(y):\nn\cr
&-&<0|\psi^\beta(x)\psi^\alpha(y)|0>:\psi^\alpha(x)\psi^\beta(y):
+<0|\psi^\alpha(x)\psi^\alpha(y)|0><0|\psi^\beta(x)\psi^\beta(y)|0>\nn\cr
&-&<0|\psi^\beta(x)\psi^\alpha(y)|0><0|\psi^\alpha(x)\psi^\beta(y)|0>
\eea
Note that when we take the 
large-$N_c$ limit we can expand the
full normal ordering in the leading order to  get 
$:\psi^\alpha(x)\psi^\alpha(y)::\psi^\beta(x)\psi^\beta(y):$.
In the above equality  the  fourth and fifth terms on the 
right  are of 
smaller order in the large-$N_c$ limit as well as the last term in the
equality. The sixth term is an infinite vacuum expectation 
value, but that is
a constant term which will not contribute to the 
equations of motion hence we can drop it. As a result, 
\bea
:\psi^\alpha(x)\psi^\beta(x)\!\!\!&::&\!\!\!\psi^\beta(y)\psi^\alpha(y):=
:\psi^\alpha(x)\psi^\alpha(y)::\psi^\beta(x)\psi^\beta(y):\nn\cr
&+&<0|\psi^\alpha(x)\psi^\alpha(y)|0>:\psi^\beta(x)\psi^\beta(y):
+<0|\psi^\beta(x)\psi^\beta(y)|0>:\psi^\alpha(x)\psi^\alpha(y):
\eea
Using the above formulae   we get a 
finite renormalization of the mass term.

Let us compute the equations of motion at the linear approximation.
What we mean by this is to  linearize the constraint as well as 
the equations of motion. 
The linearization of the constraint simply says
that 
$R(u,v)=0$ if $u,v$ have different signs.
We thus restrict ourselves to $u,v>0$ and 
compute 
\beq
    {\partial R(u,v;x^+)\over \partial x^+}=\{  R(u,v;x^+); H \},
\eeq
We also put $P=u+v, x=u/P$ and make the ansatz 
$R(u,v;x^+)=\zeta_R(x)e^{-iP_+x^+}$.
For further details we refer to the previous works 
\cite{2dqhd, istlect} where
similar calculations are done in more detail  
with  the same type of ansatz; 
this  yields  an eigenvalue 
equation,
\beq
     \mu^2\zeta_R(x)=(m^2-{g^2\over 2\pi})\Big[{1\over x}+{1\over 1-x}\Big]
\zeta_R(x)-{g^2\over 8\pi}\int_0^1 dy{\zeta_R(y)-\zeta_R(1-y)\over (y-x)^2}
,\eeq     
where $\mu^2=2P_+P$ is the invariant mass of the excitation.
By looking at the behaviour of this  equation under 
$x\mapsto 1-x$, and $y\mapsto 1-y$, we see that  
we can choose our wave functions to be   
antisymmetric under $y\mapsto 1-y$, thus $\zeta(1-y)=-\zeta(y)$.
This gives us,
\beq
  \mu^2\zeta_R(x)=(m^2-{g^2\over 2\pi})\Big[{1\over x}+{1\over 1-x}\Big]
\zeta_R(x)-{g^2\over 4\pi}\int_0^1 dy {\zeta_R(y)\over (y-x)^2}
.\eeq
This equation is one of 
our main results and it is a variant  of the well-known `t Hooft equation. 
Apart from the numerical factors this equation is 
the same as the original one, and this result fits 
to the ideas in \cite{kutasov}.
Its properties are well known, the most important one 
is that there are only bound state solutions.

An interesting  question is the existence of ``baryon" like excitations.
These should correspond to operators of the form
\beq
     {1\over Z}\epsilon_{\alpha_1\alpha_2...\alpha_{N_c}}
\chi^{\alpha_1\dag}(p_1)\chi^{\alpha_2\dag}(p_2)...
\chi^{\alpha_{N_c}\dag}(p_{N_c})
,\eeq
but the meaning of these 
operators as  $N_c\to\infty$ is not so obvious.
Yet we can think about normalized states of this form 
when {\it all the momenta are positive} acting on the 
Fock vacuum, they should correspond such baryon like 
states. Perhaps our large-$N_c$ theory can 
detect their presence. Indeed,  one can check that the 
operator,
\beq 
    \hat B={1\over N_c} \int_0^\infty [dp]\chi^{\alpha\dag}(p)\chi^\alpha(p)
,\eeq
measures the number of such excitations.
This operator can be 
given a meaning in our theory: in the large-$N_c$ limit 
therefore it is natural to expect that 
the operator, $B={1\over 2}\int_0^\infty[dp] F(p,p)$ gives us this 
number and as we will see it is well-defined.
In our classical limit we can ask if this number makes sense for our system, 
that is if it is a conserved quantity. The answer, not surprisingly,  
is no: the above baryon number {\it is not conserved} by our equations of 
motion. Thus there are really no baryons in this theory.

\section{Geometry of the Phase space}

To understand the geometry behind the  classical system
that we introduce in the previous section, we must 
take a look at the finite dimensional orthogonal group. 
Our approach will be similar to one  in 
\cite{erdal} where we discussed the bosonic version of this 
theory.
The basic ideas of the quantization  of free  Weyl
fermions and the underlying geometry is 
discussed  in the paper of Bowick and 
Rajeev \cite{bowraj}, but we would like to expand on it
and there are some differences in our conventions.

We recall  that the real orthogonal group can be defined as
the set of linear transformations which leave a quadratic form invariant.
\beq
      Q(Au,Av)=Q(u,v)
,\eeq
(here $Q(u,v)=u^TQv$ represents this quadratic form, and 
superscript  $T$ denotes the ordinary transpose).
In our case the quadratic form is diagonal, so it is the 
standard inner product $u^Tv$.
We work with  the complexification of the original real Hilbert space,
and if our Hilbert space is 
even dimensional, in this complex  
space we can use a different quadratic form,
simply by using an invertible transformation $S$,
$Q_2=S^TQS$. Assume now  that we 
have a complex structure $J$ acting on our 
original real Hilbert space, that is, a real 
antisymmetric matrix with respect to this form, which is also 
orthogonal, implying $J^2=-1$. If the quadratic form 
is the identity, we may think of such a matrix as 
$J=\pmatrix{0&1\cr -1&0}$ in an appropriate basis
of the {\it real Hilbert space}. 
Let us split our Hilbert space 
into two isomorphic pieces with respect to the 
above decomposition of the complex structure,  $W\oplus \tilde W$,  
and  complexify the real Hilbert space, naturally we have 
$W\otimes {\bf C}\oplus \tilde W\otimes {\bf C}$.  Choose
with respect to this decomposition,   
\beq 
   S=\pmatrix{ i1 & -i1\cr 1 & 1} 
.\eeq
This is the transform which we can use to diagonalize 
our complex structure.
Of course  our original quadratic form 
now changes  as we described above:  we get, 
\beq 
     Q=\pmatrix{0&1\cr 1&0}
.\eeq
(In our problem we actually transform  the inverse of this 
form, but one can see that as matrices these two forms are 
identical).
The complex orthogonal group is the set of 
transformations which leaves the form $Q$ invariant.  
Thus a general complex matrix $g=\pmatrix{a&b\cr c&d}$   is orthogonal if,
\beq
a^Tc=-c^Ta\quad a^Td+c^Tb=1\quad b^Td=-d^Tb
.\eeq 
In finite dimensions the quadratic form  is  
$Q(z,z)=z_1z_{m+1}+z_2z_{m+2}+...+z_mz_{2m}$.
We see then that the original  {\it real orthogonal group} is
embeded into the complex orthogonal group defined by 
this quadratic form as a set of matrices 
\beq
     g=\pmatrix{a&b\cr \bar b&\bar a}
,\eeq
with now $a,b$ satisfying,
$a^T\bar a+b^\dag b=1$ and $a^T\bar b=-b^\dag a$ (where 
we decomposed the matrix in the obvious way).
This explicitly shows  that the complex structure, which is a real
orthogonal matrix, becomes diagonal,
$J=\pmatrix{-i1&0\cr 0&i1}$.
In our physical example these diagonalizations will
be accomplished by the Fourier transform.

An  immediate consequence of this 
way of looking at the real orthogonal group  is that the real orthogonal 
group actually carries a copy of the unitary group in it,
corresponding to the elements,
$\pmatrix{a&0\cr 0&\bar a}$.
The quadratic form implies $a^T\bar a=1$, as well as $aa^\dag =1$,
this implies  $aa^\dag=a^\dag a=1$. 
It is the unitary group of ${\cal H}_+$, where 
${\cal H}_+$ refers to the subspace on which $J$ acts as 
$i$.

For our purposes we should extend these discussions 
to the infinite dimensional case. In the infinite 
dimensional one we should not use   the full orthogonal 
group but the one with a convergence condition\cite{bowraj}. This condition
is the well-known Hilbert-Schmidt condition in the 
quasi-free representations of canonical anticommutation algebra.
We will comment further on the convergence 
conditions  when we make contact with 
our system.
We define the restricted orthogonal group on the
complexified Hilbert space as follows,
\beq
     O_1({\cal H})=\{ g^TQg=Q| g=\pmatrix{a&b\cr \bar b &\bar a}\quad 
  b\in {\cal I}_2\}
,\eeq
where ${\cal I}_2$ is the ideal of Hilbert-Schmidt operators
\cite{simon}.
We can state the convergence condition  more economically as 
$[\epsilon, g]\in {\cal I}_2$, where $\epsilon=\pmatrix{1&0\cr 0&-1}$ with 
respect to the above decomposition.
This is basically the complex structure we had, except that a 
factor of $i$ has been removed.
The Lie algebra of this group can be  found from an
infinitesimal group element,
\beq
     g=1+i\Delta u=1+i\Delta \pmatrix{S&R\cr -\bar R&-\bar S}
,\eeq
with $R^T=-R$ and $S^\dag=S$ and $\Delta$ represents an infinitesimal 
parameter.
The reader can  verify that $u^TQ+Qu=0$.
We would like to define a classical phase space using this 
infinite dimensional orthogonal group. This 
will be our phase space for the large-$N_c$ theory, but for the 
moment let us define it as a mathematical system.
We introduce a variable $\Phi$,
\beq 
   \Phi=g \epsilon g^{-1}\quad g\in O_1({\cal H})
.\eeq
The orbit of $\epsilon$ under the restriced 
orthogonal group is parametrized by this operator.
It is easy to see that the orbit is diffeomorphic to 
\beq
    O_1({\cal H})/U({\cal H}_+)
.\eeq
The operator $\Phi$ satisfies,
\beq
    \Phi^2=1 \qquad \Phi=-Q^{-1}\Phi^T Q \qquad \Phi-\epsilon=
\pmatrix{{\cal I}_1&{\cal I}_2\cr {\cal I}_2& {\cal I}_1}
\eeq 
where ${\cal I}_1$ denotes the trace class operators in the appropriate 
space of operators( here $Q^{-1}$ is identical to $Q$ as a matrix,
but transforms differently). The second condition 
really says that $\Phi$ is in the Lie algebra of this group
(it is possible to think of this space as a real subset of 
the restricted Grassmanian, and there is an analogous construction of 
a line-bundle on this space, see \cite{borthwick}).
The tangent space of this orbit is given by the 
infinitesimal action of the group at any point, and 
in fact it is a copy of the Lie algebra of this 
group at every point. 
The action of  a vector field 
on the basic variable 
$\Phi$ becomes 
$V_u(\Phi)=i[u(\Phi),\Phi]$, for a 
Lie algebra element $u(\Phi)$, which changes
differentiably  over the orbit. So a vector field at a point 
$\Phi=g\epsilon g^{-1}$ 
 comes from  a Lie algebra element $g^{-1}u(\Phi)g$. 

It is well-known \cite{sternberg} that 
such orbits in finite dimensions typically carry a 
symplectic structure.
If we formally define a two form,
\beq
    \Omega={i\over 4} \Tr \Phi d\Phi\wedge d\Phi
,\eeq
following the methods in \cite{2dqhd} we can check  
that it is closed and non-degenerate.
The form evaluated at two vector fields $V_u, V_v$ is 
given by 
\beq
    \Omega(V_u,V_v)={i\over 8}\Tr \epsilon[[\epsilon, g^{-1}u(\Phi)g],
[ \epsilon, g^{-1}v(\Phi)g]]
\eeq
which shows that it is well-defined, due to the 
Hilbert-Schmidt conditions, non-degenerate, homogeneous and 
K\"ahler.
The group action on this phase space $\Phi\mapsto g^{-1}\Phi g$ 
is actually Hamiltonian, that is there are moment 
maps which generate this action,
given by a conditional trace, $F_u=-{1\over 2}\Tr_\epsilon(\Phi-\epsilon)u$,
with $\Tr_\epsilon(A)={1\over 2}\Tr(A+\epsilon A\epsilon)$.
Just for completeness we record that 
\beq
    \{F_u, F_v\}=F_{-i[u,v]}-2\Im {\rm m}\Tr(R_1R_2^\dag)
,\eeq
if we decompose $u,v$ as above. The last term represents a 
central part and cannot be removed in this classical theory.

How does this tie up with our system? Recall that we had a symplectic form
which was $i$ times a quadratic form  $Q$, and a Hamiltonian 
for the  free theory  which is the mass part, $i$ times an antisymmetric 
form $\omega$, the combination of the two provides  a natural operator:
$\tilde \omega=Q^{-1}\omega$ is a type $(1,1)$ tensor hence a proper linear 
transformation.  Its polar decomposition will 
have all the basic pieces we need.
 Of course we have also $\omega^{-1}Q$, so which 
one we choose is determined by the equations of 
motion. If we look at this general system in the Hamiltonian 
formalism,
\beq
   S_0 = \int dt {1\over 2}i\psi Q\partial_t \psi-\int dt H=
\int dt  {1\over 2}i \psi Q \partial_t\psi-\int dt {1\over 2}i\psi \omega \psi 
,\eeq      
the equations of motion will give us,
\beq
    \partial_t \psi=Q^{-1}\omega \psi
.\eeq
Hence the  operator $Q^{-1}\omega$ is the one we should use.
We find the polar 
decomposition of this operator, $\tilde\omega=KJ$, where $K$ is positive 
symmetric and $J^TJ=1$, orthogonal(we should be using the 
natural inner product defined by $Q$ to define the 
transpose, and in the infinite dimensional case to 
define underlying  real Hilbert space of initial data). 
However $\tilde \omega$ is antisymmetric with respect to 
our quadratic form, this means that 
$J^2=-1$  and orthogonal, thus a complex structure
( the complex structure 
coming from the other  choice  
differs from this by a minus sign).
In our example we see that the 
quadratic form is 
$2\sqrt{2}\delta(x^--y^-)\delta_{\alpha\beta}$ 
(thus all the calculations can be 
done with the usual matrix transpose),
and the antisymmetric form is $-{\sqrt{2}m^2}\partial_-^{-1}$,
so we get from the polar decomposition,
$K={m^2\over 2}[-\partial_-^2]^{-1/2}$,
$J=-[-\partial_-^2]^{1/2}\partial_-^{-1}$
(we omit the identity in the color space).
When we use a basis which diagonalizes $\tilde \omega$ we get 
solutions which oscillate in time 
with a frequency given by the eigenvalues of $K$.
In our example,
if we decompose the field $\psi^\alpha$ using a Fourier 
mode decomposition,
\beq
  \psi^\alpha(x^-)=\int_{-\infty}^\infty 
{[dp]\over 2^{3/4}} w^\alpha(p)e^{-ipx^-}
,\eeq
we have 
\beq   
    w^\alpha(p,x^+)=w^\alpha(p,0)e^{-i{m^2\over 2|p|}x^+}
{\rm\ for\ } p> 0\qquad
  w^\alpha(p,x^+)=w^\alpha(p,0)e^{+i{m^2\over 2|p|}x^+}{\rm \ for\ } p<0
.\eeq
(Note that the above combinations on the exponents are 
relativistically invariant if we recall the mass-shell condition
$p_+={m^2\over 2 p_-}$).
This suggests that the $i$ subspace of $J$ goes to 
creation operators, and $-i$ subspace goes to 
the annihilation operators, it is better therefore to represent our 
Fourier coefficients as  
$w^\alpha(p)=\xi^\alpha(p)$  and $w^\alpha(-p)=\bar \xi^\alpha(p)$  for $p>0$.
If we act with $J$ on our 
field variables,
\beq
      (J\psi)^\alpha(x^-)=\int_0^\infty {[dp]\over 2^{3/4}}
(-i\bar \xi^\alpha(p) e^{-ipx^-}+i\xi^{\alpha}(p)e^{ipx^-})
.\eeq
We see now that this Fourier transform diagonalizes our 
complex structure. If we look at the inverse of the quadratic form 
it transforms as 
$\int dx^-dy^-2^{3/4}e^{ipx^-}(2\sqrt{2})^{-1}\delta(x^--y^-)2^{3/4}e^{iqy^-}$,
which gives us $\delta[p+q]$. This is the form of $Q$ that 
we wanted to obtain.

From the Fourier decomposition, 
creation and annihilation operators therefore are 
assigned according to $\sgn(p)$, $\xi(p)\mapsto \chi^{\dag\alpha}(p)$ and
$\bar \xi(p)\mapsto \chi^{\alpha}(p)$.
The ultimate reason for the choice of Fock vacuum is to make 
the Hamiltonian bounded from below,
if we write our Hamiltonian in the Fourier space,
\bea
    H_0={\cal P}\int {[dp]\over 2^{3/2}}
  {\sqrt{2}m^2\over 2|p|}\sgn(p):\chi^\alpha(-p)\chi^\alpha(p):=
\int_{0^+}^\infty {[dp]\over 2^{3/2}}{\sqrt{2}m^2\over 2|p|}
[:\chi^{\alpha\dag}(p)\chi^\alpha(p):-:\chi^\alpha(-p)
\chi^{\alpha\dag}(-p):]\nn
.\eea
Notice that sgn$(p)$ appears in the Hamiltonian, which is 
basically the complex structure we have, and the normal ordering (according to 
our choice of creation and annihilation operators) now makes the Hamiltonian 
bounded from below:
\beq
    H_0=\int_{0^+}^\infty {m^2\over 2|p|}\chi^{\alpha \dag}(p)\chi^\alpha(p)
.\eeq
We could question the effect of the interactions since we have been 
describing everything in terms of the free part of the Hamiltonian.
Here we see a clear advantage of our light-cone point of view,
the complex sturcture we start with using the free Hamiltonian is
independent of any of the parameters of the theory, {\it thus 
the choice of quasi-free  representation of the canonical anticommutation
relations is not affected by the change of parameters due to interactions}.
In our case we explicitly keep the change of mass due to the 
interactions with the gauge
fields, so we are not taking advantage of this property.
In more general case this property may be helpful, in fact for the
scalar theory it is essential. 
We thus conclude our discussion on the choice of Fock space and 
its relation to the natural complex structure in our system.

Next  we  show that $\Phi-\epsilon$  really represents our basic bilinears:
let us decompose the complexification of  our one-particle Hilbert space 
as ${\cal H}_+\oplus{\cal H}_-$ according to  
$-\sgn(p)$, we can write a general bilinear as an operator acting on the 
one-particle space and decomposed according to this 
direct sum, 
one checks that 
\beq 
      F=\pmatrix{ S&R\cr -\bar R &-\bar S}
,\eeq 
with exactly $S^\dag=S$ and $R^T=-R$.
We also know that $(F+\epsilon)^2=1$. But these are 
exactly the properties satisfied by $\Phi$.
Our physical system has a one-particle Hilbert space 
given by the initial data on the light-cone $x^+=0$,
we complexify this 
space and use Fourier transform to put 
our operators into the desired form.
Then ${\cal H}_-$ corresponds to the negative 
frequency components in the physics language. 
The Poisson bracket relations can be meaningfully extended to
the Hilbert-Schmidt type $R$, so we need the 
convergence conditions.   
The convergence conditions are also a natural consequence of the
super-renormalizability of this system. 
The time evolution of the finite $N_c$ system should keep us in the 
same free Fock space, and in the large-$N_c$ limit 
this should be expressible as an operator like $\Phi$.
In fact the smeared out Poisson brackets are given by
the Poisson bracket relations of the moment maps. 
Thus 
the symplectic structure we have on this homogeneous manifold 
is the one we have found for our bilinears.

It is useful to look at the same issue 
 from the point of view of generalized coherent states:
assume that we have a Lie group which is representable  
on a Hilbert space by unitary operators through a 
highest weight vector. If we look at the orbit of this 
vector under the action of the group, this 
orbit has a natural symplectic structure, and 
all the vectors on the orbit correspond to 
the generalized coherent states \cite{perelomov, onofri, yaffe}.
In our case the group of Bogoluibov automorphisms, 
which do not act on the color part of
our fermions
are  represented on the fermion Fock space by the 
color invariant bilinears.  The highest weight vector is 
the vacuum and its orbit under this group therefore carries a 
natural symplectic structure. 
The corresponding  group is the restricted orthogonal 
group $O_1({\cal H})$ and the orbit 
is our phase space.(In fact physically 
we should be using the projective Fock space, since the phase
does not change the physical content of a state.
The bilinears provide a unitary representation of the
central extension $\hat O_1({\cal H})$ of the group 
$O_1({\cal H})$, when we use the projective Fock space, 
the central part disappears and we decend to
the restricted orthogonal group). 
 The convergence 
conditions are now a result of the implementability 
of these automorphism in the Fock space, which is 
defined by our choice of the vacuum \cite{araki, mickbook, pressegal}.
The large-$N_c$ limit allows us to restrict to
the bilinears and the super-normalizability 
keeps us in the restricted class of implementable 
automorphisms. Thus taking the large-$N_c$ 
limit provides  a classical limit in this sense.

This shows that our large-$N_c$ limit 
has a well-defined classical phase space with a natural 
symplectic structure. This openes up various possibilities, such as 
studying large fluctuations of the field in this limit. There are various
delicate questions, such as the domain of the Hamiltonian, existence 
of finite time evolution, completeness of the trajectories which we 
plan to come back in the future.

\section{Bosons and Fermions}

This is the begining of the second part of our paper.
The second part has  two themes again: the construction of the 
phase space via the large-$N_c$ limits of 
the bilinears and the geometry of the 
ensuing  phase space.  Since the bosonic theory is
developed in \cite{erdal} and the fermionic version 
is explained in detail in the previous sections
the construction of the phase space and finding the Hamiltonian 
will be very brief. We recommend  the reader 
to look at \cite{erdal} and we use the results of 
the previous sections freely. 
The geometry part, which is in the 
next section,  will require new methods and 
in some sense it is not as complete.
It may be helpful if the reader also consults 
to \cite{tolyateo} where the $SU(N_c)$ version is
discussed.
We will develop these aspects as much as we can and in some 
cases we indicate what the idea should  be. 

We start  our first theme: 
 we  use the same conventions as in the previous sections and 
 our previous paper.
The action functional of the combined system 
of bosons and fermions can be 
written as,
\bea
    S=\int \Big[ {1\over 2}\Tr F_{\mu\nu}F^{\mu\nu}+
i\bar \Psi_M \gamma^\mu D_{\mu} \Psi_M -m_F\bar \Psi_M\Psi_M+
{1\over 2} (D^\mu \phi)^T(D_\mu \phi)-{1\over 2}m_B^2 \phi^T\phi\Big]
,\eea
where we use the same conventions as in section IV for the 
Majorana fermions. The transpose refers to the color indices 
for the scalar field. Again the covariant derivative is $D_\mu=\partial_\mu+
gA_\mu$, where $A_\mu$ has values in the Lie algebra of $SO(N_c)$.  
We choose $x^+$ as time and set $A_-=0$
as our gauge fixing condition. 
Then the action in the light-cone formalism reads, 
\bea
 S&=&\int dx^+dx^-\Big[i\sqrt{2} \psi_1^T
\partial_-\psi_1+i\sqrt{2}\psi_2^T\partial_+\psi_2+2im_F\psi_1^T\psi_2\nn\cr
&\ &+{1\over 2}\phi^T(-2\partial_-)\partial_+\phi-{m_B^2\over 2}\phi^T\phi
+gA_+^a[i\sqrt{2}\psi_2^TT^a\psi_2+{1\over 2}(\partial_-\phi^TT^a\phi-
\phi^TT^a\partial_-\phi)]+{1\over 2}(\partial_-A_+^a)^2\Big]
.\eea
The advantage of the light-cone formalism is again  clear, we 
are already in the Hamiltonian picture. We can read off the Poisson brackets 
satisfied by the dynamical fields.
We also see  that  $\psi_1$ is not dynamical, 
as well as $A^a_+$,
therefore they  can be eliminated through their equations
of motion. 
The dynamical fermion field 
$\psi_2^\alpha$ will be called $\psi^\alpha$ for 
simplicity as in the previous sections. We will assume that  the field 
$A^a_+$ is eliminated after the dynamical 
fields are quantized, this will give us
the quantized  Hamiltonian of the system,
\beq
   H=\int dx^-\Big({1\over 2} m_B^2:\phi^T \phi:+{1\over 2} \sqrt{2}m_F^2
:\psi^T{1\over i\partial_-}\psi: -{g^2\over 2}:J^a: {1\over \partial_-^2}:J^a:
\Big)
,\eeq
where
\beq 
    J^a=\Big[i\sqrt{2}\psi^T T^a\psi +{1\over 2} (\partial_-\phi^TT^a\phi
-\phi^TT^a\partial_-\phi)\Big]
.\eeq
The quantization process is defined for the 
Fermionic sector in section II and for bosons  in 
the reference \cite{erdal}.
We expand fermions and bosons into Fourier modes 
in a {\it complex} space,
\beq
     \psi^\alpha(x^-)=\int_{-\infty}^{\infty} {[dp]\over 2^{3/4}}\chi^\alpha(p)
e^{-ipx^-} \quad \phi^\alpha(x^-)=\int_{-\infty}^{\infty} 
{[dp] \over \sqrt{2|p|}} a^\alpha(p)e^{-ipx^-}
,\eeq
with now 
$\chi^{\alpha(p)\dag}=\chi^\alpha(-p)$ and 
$a^{\alpha\dag}(p)=a^\alpha(-p)$. 
(We should again assume that there is an infinitesimal cut-off 
around the zero momentum to be taken to zero at a later stage).
The Poisson bracket relations 
go to 
\beq
    [\chi^\alpha(p),\chi^\beta(q)]_+=\delta[p+q]\delta^{\alpha\beta},\quad 
[a^\alpha(p), a^\beta(q)]=\sgn(p)\delta[p+q]\delta^{\alpha\beta}.
\eeq
These are exactly the same as before, 
there is one more commutator now,
\beq
   [\chi^\alpha(p), a^\beta(q)]=0
.\eeq
As we will 
see in the next section  the definition of the 
Fock vacuum brings new features-a larger symmetry algebra 
appears.
We introduce  the vacuum state $|0>_s$, characterized by 
$\chi^\alpha(p)|0>_s=0, a^\alpha(p)|0>_s=0$ for $p > 0$,
where we put a subscript $\ _s$ to emphasize that the 
vacuum is for the full algebra of the boson/fermion system.
We repeat for the convenience of the 
reader the normal ordering rules of the bilinears
(rewritten to fit to our needs),
\bea
  :\chi^\alpha(p)\chi^\beta(q):&=&\chi^\alpha(p)\chi^\beta(q)-
{1\over 2} \delta^{\alpha\beta} (1+\sgn(p))\delta[p+q]\nn\cr
  :a^\alpha(p)a^\beta(q):&=& a^\alpha(p)a^\beta (q)-{1\over 2}
\delta^{\alpha\beta} (1+\sgn(p))\delta[p+q].
\eea
There is an obvious extension of the general definition of normal 
ordering  to the 
product of more than two  operators, which one needs 
for the reduction of the Hamiltonian: set all the annihilation 
operators to the right of creation operators in a recursive way.

We first introduce our bilinears for the large-$N_c$ limit 
and work out their Poisson brackets.
Then we express our Hamiltonian in the large-$N_c$ limit 
in terms of these bilinears.
We can see that the basic color invariant observables are:
\bea
&\ & \hat F(p,q)={2\over N_c} :\chi^{\alpha\dag}(p) \chi^\alpha(q): \quad 
\hat     B(p,q)={2\over N_c} :a^{\alpha\dag}(p) a^\alpha(q):\nn\cr 
&\ & \hat C(p,q)={2\over N_c} \chi^{\alpha\dag}(p)  a^\alpha(q) \quad
\ \ \hat {\bar C}(p,q)={2\over N_c} a^{\alpha\dag}  (p)\chi^\alpha (q) 
\eea
note that we have no need for normal ordering in the 
last two operators since they consist of commuting  operators. 
In the large-$N_c$  limit $\hat C$ and  $\hat {\bar C}$ are related,
$\bar C= C^\dag$, and there are 
similar conditions on $F,B$ (when we represent 
the resulting classical observables as integral kernels and 
think of them as  now abstract operators).
For our calculational purposes it is better to introduce the 
following variables as in section III and the reference \cite{erdal},
\beq 
    \hat T(p,q)={2\over N_c}:a^\alpha(p)a^\alpha(q):=\hat B(-p,q)\quad
  \hat  R(p,q)={2\over N_c} :\chi^\alpha(p)\chi^\alpha(q):=\hat F(-p,q)
,\eeq
and also the  variable,
\beq 
\hat S(p,q)={2\over N_c} \chi^\alpha(p)a^\alpha(q)=\hat C(-p,q)
.\eeq
These variables in the large-$N_c$ limit satisfy the following (super)Poisson 
brackets,
\bea
   \{ T(p,q),  T(s,t)\}&=&
 -2i\Big( {\rm sgn} (p) \delta[p+s]  T(q,t)
+{\rm sgn}(q)\delta[q+s] T(p,t)+{\rm sgn}(p)\delta[p+t] T(s,q)\nn\cr
&+&{\rm sgn}(q)\delta[q+t] T(s,p)
 + ({\rm sgn} (p)+{\rm sgn} (q) )(\delta[p+s]\delta[q+t] 
+\delta[p+t]\delta[s+q])\Big)\nn\cr
\{ R(p,q), R(s,t) \}&=&-2i\Big(R(p,t)\delta[q+s]-R(q,t)\delta[p+s]
+R(s,p)\delta[q+t]-R(s,q)\delta[p+t]\nn\cr
&+&(\delta[q+t]\delta[p+s]-\delta[p+t]\delta[q+s])({\rm sgn}(t)+
{\rm sgn}(s))\Big)\nn\cr  
\{T(p,q), S(s,t)\}&=& -2i\Big( S(s,q)\sgn(p)\delta[p+t]+
S(s,p)\sgn(q)\delta[q+t]\Big)\nn\cr
  \{R(p,q), S(s,t)\}&=& -2i\Big(
  S(p,t)\delta[q+s]-S(q,t)\delta[p+s]\Big)\nn\cr
 \{ S(p,q), S(s,t)\}_+&=&-2i\Big(T(q,t)\delta[p+s]-R(s,p)\sgn(q)\delta[q+t]
+\delta[p+s]\delta[q+t](1+\sgn(p)\sgn(q))\Big)
.\eea   
We note that the last one is  symmetric in the variables and 
the third and forth ones show that $S$ behaves as a 
module of the algebras defined by the 
Poisson brackets of $T,R$, thus it carries a representation of 
these two algebras. This is the general form of a 
super-algebra structure.
We will denote the full set of these 
brackects as a super-Possion bracket $\{\ , \ \}_s$.

The conversion  of the 
normal ordered products of non-color invariant 
combinations appearing in the above Hamiltonian
to the full normal ordering in the 
large-$N_c$ limit can be achieved as before resulting with the
same changes in the masses $m_F^2\mapsto m_F^2-g^2/2\pi$ and
$m_B^2\mapsto m_R^2-g^2/2\pi$,
where $m_R^2=m_B^2-g^2/4\pi \ln(\Lambda_U/\Lambda_I)$ denotes the 
renormalized mass of the boson.  
We skip the details of this reduction, since 
they are the extensions of the details in \cite{istlect} and we
have given some essential steps in section III.
The resulting Hamiltonian of our system in the 
large-$N_c$ can be expressed as a free part and an interacting part:
\beq 
    H_0={1\over 8}(m_R^2-{g^2\over 2\pi}){\cal P}\int {[dp]\over |p|}
T(-p,p)+{1\over 8}(m_F^2-{g^2\over 2\pi}){\cal P}\int {[dp]\over p}
R(-p,p)
.\eeq
The interaction part is written as
\bea
H_I&=&{\cal FP}\int [dpdqdsdt]\Big( G_1(p,q;s,t)T(p,q)T(s,t)+
G_2(p,q;s,t)R(p,q)R(s,t)\nn\cr
  &\  & +G_3(p,q;s,t)S(p,q)S(s,t)\Big),
\eea
where the kernels are given by 
\bea
    G_1(p,q;s,t)&=&{g^2\over 64}{\delta[p+q+s+t]\over \sqrt{|pqst|}}
{sq-st+pt-pq\over (p+s)^2}\nn\cr
  G_2(p,q;s,t)&=& -{g^2\over 64}{\delta[p+q+s+t]\over (p+s)^2}\nn\cr
  G_3(p,q;s,t)&=& {g^2\over 64}{\delta[p+q+s+t]\over \sqrt{|tq|}(p+s)^2}
.\eea

We have  not completed the definition of our large-$N_c$ limit yet,
there is a constraint. Recall that we still have a 
left over global color invariance, which is generated by the
operator,
\beq
\hat  Q^{\alpha\beta}= \int [dp](:\chi^{\alpha\dag} (p)\chi^\beta(p):  
+\sgn(p):a^{\alpha\dag} (p)a^\beta(p): )
.\eeq
When we restrict our color invariant 
bilinears to the color invariant sector of the 
full Fock space, we find that 
\bea
&\ &(F+\epsilon)^2+C\epsilon C^\dag=1\nn\cr
&\ &C\epsilon B+C\epsilon +FC+\epsilon C=0\nn\cr
&\ & \epsilon B \epsilon C^\dag +C^\dag +\epsilon C^\dag \epsilon+
\epsilon C^\dag F=0\nn\cr
&\ & (\epsilon B +\epsilon)^2+\epsilon C^\dag C=1
,\eea 
where we  define  as in section III, 
$\epsilon(p,q)=-\sgn(p)\delta[p-q]$
(here the minus sign is crucial, in our previous works that was not 
important, but in the super case there is a prefered choice) and 
we also employ the product convention as before for example $(FC)(p,s)=
\int [dq]F(p,q)C(q,s)$. We warn the reader that above the 
two epsilons have the same matrix elements but they are 
acting on different spaces. 
The meaning of this constraint could best be understood if we 
introduce a  super operator,
\beq
\Phi=\pmatrix{\epsilon B +\epsilon&  \epsilon C^\dag \cr  C& F +\epsilon}      
.\eeq
The above  constraint is simply given by 
\beq 
\Phi^2=1
.\eeq
It also satisfies  a Lie algebra condition, it is 
better to write it in the following form:
use a decomposition of our super-space into 
${\cal H}_+|{\cal H}_+\oplus {\cal H}_-|{\cal H}_-$,
according to the sign of $\epsilon$ in even and odd parts respectively.
Then we have $\hat \epsilon=\pmatrix{1&0\cr 0&-1}$, 
and we introduce with respect to this 
decomposition $\hat \omega_s=\pmatrix{0&-\bar \epsilon\cr 1&0}$, then:
\beq
\hat \omega_s \Phi^\tau+\Phi\hat \omega_s=0
,\eeq
we invite the reader to verify this.

There are also convergence conditions, which come from the 
super-renormalizability of this system again. The time evolution should 
leave this system in the same Fock space.
Another way to see this is to think about the 
smeared out operators, and see that the 
central terms make sense only for the restricted set of
operators, for which the off-diagonal blocks are in the 
Hilbert-Schmidt class.
We can write down these convergence conditions in an economical way as
\beq
    [\hat \epsilon, \Phi] \in {\cal I}_2
,\eeq
where ${\cal I}_2$ refers to the ideal of
Hilbert-Schmidt operators in this super-space.
We have proposed elsewhere \cite{teoman} a method of 
introducing such operators in the super-context, and we assume 
this definition is used. Since these technical matters 
are not completely settled we are brief at this point, see also 
the next section on the geometry. 
This completes the construction of our large-$N_c$ limit: 
we  postulate the above Hamiltonian, the 
super-Possion brackets with the constraint and 
this defines a classical system. 
The ``time'' evolution  is given by the basic rule:
for any observable $O_s$ of the theory
\beq 
     {\partial O_s\over \partial x^+}=\{ O_s, H\}_s
,\eeq 
where the Hamiltonian is in general 
an even function of our bilinears--
which we should consider as the coordinates of this 
phase space.

It is possible to carry out the analysis given in 
\cite{tolyateo}, but we will be content with describing only 
the linear approximation. We plan to report on these 
in a separate publication (they will appear in the 
PhD thesis of the first author).
 
We start with the linearization of the constraint $\Phi^2=1$, which 
gives us 
\beq
F\epsilon+\epsilon F=0\quad \epsilon B \epsilon+B=0\quad \epsilon C+C \epsilon
=0.
\eeq
The first two are exactly the conditions we
 have found before, for mesons made up of only bosons in \cite{erdal},
the first one is in section III, and the last one is the 
new condition on our odd variable.
In terms of $S$ that means we have 
$S(u,v)=0$ unless $u,v>0$ or $u,v<0$.
If we assume $u,v>0$ and evaluate the equations of 
motion in the linear approximation for $S(u,v)$,
$\partial_+S(u,v;x^+)=\{ S(u,v;x^+), H\}$ and 
furthermore we make the same type of ansatz as in 
\cite{istlect, erdal, tolyateo} $S(u,v;x^+)=\zeta_S(x)e^{-iP_+x^+}$,
with $P=u+v, x={u\over P}$, 
\beq
    \mu_s^2\zeta_S(x)=[{m_F^2-g^2/ 2\pi\over x}+
{m_R^2-g^2/ 2\pi\over 1-x}]\zeta_S(x)-{g^2\over 8\pi}
\int_0^1 {dy\over \sqrt{xy}}{x+y\over (x-y)^2}\zeta_S(1-y)
.\eeq
The other linearized equations are the same as before
(see section III  and \cite{erdal}).

There are baryonic states that we can measure by the 
operator 
\beq 
    {\bf \hat B}={1\over N_c}\int_0^\infty [dp] 
(\chi^{\alpha\dag}(p)\chi^\alpha(p)
+a^{\alpha \dag}(p)a^\alpha(p) )
,\eeq
in the large-$N_c$ limit this 
operator should go to ${\bf B}={1\over 2}\int_0^\infty[dp] ( F(p,p)+
B(p,p))$.
The {\it baryonic states} for finite $N_c$  correspond to 
states of the form 
\bea
{1\over Z} \epsilon_{\alpha_1\alpha_2...\alpha_{N_c}}
\chi^{\alpha_1\dag}(p_1)...\chi^{\alpha_s\dag}(p_s)
a^{\alpha_{s+1}\dag}(p_{s+1})...a^{\alpha_{N_c}\dag}(p_{N_c})
,\eea
where $p_1...p_{N_c}>0$ and 
products of them acting on $ |0>_s$.

Not surprisingly {\it the above baryon number is not a conserved
quantity},  
so it does not have the physical importance as it has 
in the case of Dirac fermions where it is a conserved number,
in fact a topological number(see \cite{istlect} for the 
discussion of this in the large-$N_c$ limit and its extension in 
\cite{tolyateo}).

\section{The Geometry of the Phase Space }

Let us define a super space ${\cal H}|{\cal H}$, where we 
use a splitting   to  even and odd according to the grading $+,-$
(we are using a ${\bf Z}_2$ graded {\it real Hilbert space}).
We recall some of the conventions, following Berezin \cite{berezin}:
we work with the Grassmann envelop of this graded vector space
(thus we acquire a {\bf Z} grading).
Its mathematical theory is delicate and we will comment on it later
(some good examples of homogeneous  super-symplectic manifolds are
worked out in \cite{borthwick1}, this is a good reference to learn by 
examples).
We decompose every linear transformation or tensor according to this 
grading, the standard matrix form of a linear transformation is
\beq
    \pmatrix{A&B\cr C&D}:{\cal H}|{\cal H} \to {\cal H}|{\cal H}
,\eeq
where $A,D$ are even and $B,C$ are odd.
This means that $A=A_B+A_S,D=D_B+D_S$ where subscript $\ _B$ refers to 
the body that is 
the ordinary numbers, subsrcipt $\ _S$ refers to 
the soul, that is only the Grassmann part. $B,C$ have no body they are purely 
Grassmann valued.

We have the usual hermitian conjugation of such 
block matrices, but the transpose has to be carefully defined.
We introduce a super-transpose, $\tau$,
\beq
  \pmatrix{A&B\cr C&D}^\tau=\pmatrix{A^T&C^T\cr -B^T&D^T}
\eeq
where $T$ denotes the ordinary matrix transpose.
One can verify that this form satisfies 
$(AB)^\tau=B^\tau A^\tau$. It will be useful to record the 
following properties,
$\Str(A^\tau)=\Str A$,
if we decompose our graded space into a direct sum, for example in 
our case into ${\cal H}_+|{\cal H}_+\oplus {\cal H}_-|{\cal H}_-$,
the operators can also be decomposed into super-operators,
say into $\pmatrix{a&b\cr c&d}$, then 
\beq
    \pmatrix{a&b\cr c&d}^\tau=\pmatrix{a^\tau&c^\tau\cr b^\tau&d^\tau}
.\eeq
Realness is related to an involution in the 
Grassmann algebra,  $\xi \mapsto \xi^*$ and we assume 
that this involution obeys 
$(\xi^i\xi^j)^*=(\xi^j)^*(\xi^i)^*$ and
$(a\xi)^*=\bar a \xi^*$, where $a$ is a 
complex number and bar denotes the ordinary complex conjugation.
The real Grassmann algebra is the part which is 
invariant under this involution.
This means that there will be 
factors of $i$ to make things invariant.
This implies that {\it the real graded Hilbert space is 
defined residing inside a complex graded Hilbert space}.

On the space of linear transformations there is 
a complex conjugation operator, according to 
Berezin conventions it should be given by
the following:
write a linear transformation in its standard form,
then
\beq
    \pmatrix{a&b\cr c&d}^*=\pmatrix{a^*&-b^*\cr c^*&d^*}
.\eeq
we note that $A^{**}=A$, and 
$(A^\tau)^*=\pmatrix{a^\dag&-c^\dag\cr -b^\dag&d^\dag}=\tilde E A^\dag \tilde 
E$, here $\tilde E=\pmatrix{1&0\cr 0&-1}$, whereas $(A^*)^\tau=A^\dag$.

We have the  set  of {\it real linear transformations},
this set is invariant under the above conjugation, 
$M^*=M$, 
it remains so under the 
product of super-matrices,  
thanks to  $(A_1A_2)^*=A_1A_2$.
The set of real linear operators thus is an 
algebra.

Let us assume that the even part has a symplectic form $\omega$ and 
the odd part has a standard quadratic form $1$.
On the complexification of this space we introduce a super-symplectic form,
\beq
     \omega_s=\pmatrix{\omega &0\cr 0&i1 }, \quad \omega=\pmatrix{0&-1\cr 1&0}
.\eeq
Note that multiplying the last part with an $i$ does not really change
anything as far as only the even 
transformations are concerned but for the full case we need this 
factor.
We look at the space of {\it real transformations} which 
will leave this form invariant,
\beq
    g^\tau \omega_s g=g\omega_s g^\tau=\omega_s
,\eeq
this is a super-group, and it is denoted 
by $OSp({\cal H}|{\cal H})$.
Its even part has body isomorphic to 
$Sp({\cal H})\oplus O({\cal H})$, the odd 
parts are modules over the Grassmann envelops of these 
groups.
If we write down the group conditions 
for an element $\pmatrix{a&b\cr c&d}=\pmatrix{a^*&-b^*\cr c^*&d^*}$, 
where $a,d$ are real 
even, $c$ real odd  and $b$ imaginary odd operators,
\beq
      a^T\omega a+ic^Tc=\omega\quad a^T\omega b+ic^Td=0\quad
-b^T\omega a+id^Tc=0\quad -b^T\omega b+id^Td=i1
.\eeq
Since we have the complex conjugation convention 
$(\psi\xi)^*=-\psi^*\xi^*$, the complex conjugate of 
a product of odd operators become imaginary, this is why we have 
$ic^Tc$, then it becomes a real even element of the
Grassmann envelop.
 
Decompose our spaces according to the  
matrix representation of $\omega_s$, 
$W\oplus\tilde W|W\oplus \tilde W$.
Let us assume that we also have a super-complex structure,
which is a type $(1,1)$ tensor,
\beq 
    J_s=\pmatrix{J&0\cr 0&J}\quad J=\pmatrix{0&1\cr -1&0}
.\eeq
Assume that we extend everything to the complexification of our original
Hilbert space. The we can 
 perform a transformation 
$S$ that will put the above complex structure
into diagonal form in this complexified
space. To accomplish this it is better to represent
it in a slightly different way,
use a decomposition $W|W\oplus \tilde W|\tilde W$, then  
\beq 
  J_s=\pmatrix{0&1\cr -1&0}\quad S={1\over \sqrt{2}}\pmatrix{i&-i\cr 1&1}
\eeq
Now compute $S^{-1}J_sS$ and see that we
get $\hat J_s=i\hat \epsilon=\pmatrix{-i&0\cr 0&i}$,
which defines $\hat \epsilon$ in this decomposition.
We use a decomposition according to the sign of 
$i$  and   the resulting graded Hilbert space
becomes ${\cal H}_+|{\cal H}_+\oplus{\cal H}_-|{\cal H}_-$.
If we compute the transformation of $\omega_s$,
it goes into 
$S^\tau \omega_s S$ since it is a two form, and we get 
\beq
     \hat \omega_s = \pmatrix{0&-\bar \epsilon \cr 1 &0}, \quad \bar \epsilon=
\pmatrix{1&0\cr 0&-1},
\eeq
with respect to the above decomposition.
Obviously our real group also transformed by the same 
rule as $J_s$,
so a typical group element becomes according to the 
above decomposition,
\beq
    g=\pmatrix{A&B\cr B^*& A^*}
.\eeq
Note that each of the blocks are super-operators with standard decompositions,
and for each one 
we are using  Berezin definition of the complex conjugate $\pmatrix{a&b\cr 
c&d}^*=\pmatrix{a^*&-b^*\cr c^*&d^*}$.
We have a full complex group which leaves invariant the above 
transformed version of the  two form, this is the complex 
$OSp$ group,
\beq
     g^\tau \hat \omega_s g=\hat \omega_s, \quad g=\pmatrix{A&B\cr C&D}
,\eeq
and the real group now sits inside this complex group.
Thus  a complex transformation satisfies 
\beq
  A^\tau \bar \epsilon C-C^\tau A=0\quad 
A^\tau\bar \epsilon D-C^\tau B=\bar \epsilon\quad
-B^\tau\bar \epsilon  C+D^\tau A=1\quad B^\tau\bar\epsilon  D-D^\tau B=0
.\eeq
The reader may question the consistency of these 
equations. We should remember that 
$(M^\tau)^\tau=\bar \epsilon M\bar \epsilon$, then we can see 
that they are consistent.
There is an interesting subgroup,
given by elements of the form $g=\pmatrix{A&0\cr 0&A^*}$ and $A$ satisfies 
\beq 
    A^\tau \bar \epsilon A^*=\bar \epsilon \quad A^{*\tau} A=1
,\eeq
recall that $A^{*\tau}=A^\dag$ and $(A^\tau)^*=\bar \epsilon A^\dag \bar 
\epsilon$,
this is the same as before except that we express 
it in the subspace, so we should use $\bar \epsilon$ instead of $\tilde E$,
we get 
$A^\dag  A=A A^\dag =1$.
Let us see what it means when we expand 
$A=\pmatrix{a&\beta\cr \gamma&d}$,
\beq 
 a^\dag a=1+\gamma^\dag \gamma\quad a^\dag\beta+\gamma^\dag d=0\quad
 \beta^\dag a+d^\dag\gamma=0\quad d^\dag d=1+\beta^\dag \beta
,\eeq  
we wee that the body parts satisfy 
$a_B^\dag a_B=1, d^\dag_B d_B=1$, these are the ordinary 
unitary groups inside.
Therefore we have shown that this group's even part 
has body $U({\cal H}_+)\oplus U({\cal H}_+)$. 
This group is   denoted by 
$U({\cal H}_+|{\cal H}_+)$ and it is 
the super-unitary group of ${\cal H}_+$.

Let us define the orbit of 
$\hat \epsilon$, this  is really the complex structure if we remove the
factors of $i$, under the real group 
$OSp$:
\beq
      \Phi=g^{-1}\hat \epsilon g
.\eeq
It is immediate that $\Phi^2=1$.
We will now show that we also have 
\beq
     \hat \omega_s \Phi^\tau +\Phi \hat \omega_s=0
,\eeq
so it is an element of the Lie algebra of $OSp$.
Define $\hat E=\pmatrix{\bar \epsilon&0\cr 0&\bar \epsilon}$,
this is really our $\tilde E$ written in this splitting of the
Hilbert space, 
and note $(\hat \omega_s^\tau)^{-1}=\hat \omega_s$, 
$\hat \omega_s^\tau=\hat \omega_s \hat E$,
$\hat \omega_s^2=\hat E$ and
$\hat \epsilon\hat \omega_s=-\hat \omega_s\hat \epsilon$, then,
\bea
  \Phi^\tau \omega_s&=&(\hat \omega_s^{-1} g^\tau \hat \omega_s
\hat \epsilon g)^\tau \omega_s=g^\tau\hat \epsilon\hat \omega_s \hat E
g^{\tau\tau}\hat E=g^\tau\hat \epsilon \hat \omega_s g\nn\cr
&=&-g^\tau\hat \omega_s\hat \epsilon g=-\hat \omega_s\hat 
\omega_s^{-1} g^\tau\hat \omega_s\hat \epsilon g=-\hat \omega_s
g^{-1} \hat \epsilon g=-\hat \omega_s\Phi
,\eea
where we used $\hat E g^{\tau\tau} \hat E=g$.
Let us look at the 
stability subgroup of $\hat \epsilon$,
that is given by 
operators of the form 
$\pmatrix{A&0\cr 0&A^*}$, and we have seen that this can be 
identified with the unitary operators on ${\cal H}_+|{\cal H}_+ $, 
$U({\cal H}_+|{\cal H}_+)$.
Hence we conclude that 
our  variable $\Phi$ is actually parametrizing the space 
\beq
 OSp({\cal H}|{\cal H}) /U({\cal H}_+|{\cal H}_+)  
.\eeq
What is the advantage of this parametrization? The above super manifold 
is actually a symplectic manifold with a super-symplectic structure most
naturally written in terms of the variable $\Phi$:
\beq 
   \Omega_s={i\over 4} \Str \Phi d\Phi\wedge d\Phi
.\eeq
This is formally defined, but we use the rules of
super analysis to define our differential forms.
Clearly it is closed,
use 
\bea
     d\Str \Phi d\Phi\wedge d\Phi&=&\Str d\Phi\wedge d\Phi\wedge d\Phi=
\Str \Phi^2 d\Phi\wedge d\Phi \wedge d\Phi\nn\cr
&=& \Str \Phi d\Phi\wedge d\Phi \wedge d\Phi \Phi=
-\Str \Phi^2 d\Phi\wedge d\Phi \wedge d\Phi \wedge d\Phi.
,\eea
where we used $\Str AB= \Str BA$.
It is also clear that 
this form is homogeneous.
Its nondegenaracy can be proved at 
$\hat \epsilon$, and homogeneity proves it everywhere.

Upto now we have really used a finite dimensional approach, but  
to identify the large-$N_c$ phase space of the previous section, we need 
to extend these notions to the infinite dimensional case.
The extension is formally simple, we assume that we have 
super-Hilbert spaces, that is 
even and odd spaces each one are coming from a 
separable Hilbert space and we use a proper extension of 
the Grassmann envelop to this case(this is not so obvious and 
we assume our proposal in \cite{teoman}, 
this may not be the only possibility
see \cite{schmitt1, schmitt2}). In this infinite 
dimensional setting we introduce 
a Hilbert-Schmidt condition, 
the group that we use should be the restricted 
{\it real } $OSp$ group, 
\beq
    OSp_1({\cal H}|{\cal H})=\{ g=\pmatrix{A&B\cr B^*&A^*}| g^{-1} 
{\rm \ exists\ }, g^\tau \hat \omega_s \epsilon g=1 \quad 
[\hat \epsilon, g]\in {\cal I}_2 \}
.\eeq
The variable $\Phi$ now satisfies some convergence conditions,
indeed one can check that 
\beq
    \Phi-\hat \epsilon \in \pmatrix{{\cal I}_1 & {\cal I}_2\cr 
  {\cal I}_2 & {\cal I}_1},
\eeq
where each block refers to a super-operator in the appropriate 
class of operator ideal.
These convergence conditions imply that the 
super-symplectic form $\Omega_s$ we defined  in the 
finite dimensional setting makes sense.
The trace class conditions are important to write down 
moment maps, but we will ignore it for this work.
Hence we have an infinite dimensional phase space,
\beq 
     OSp_1({\cal H}| {\cal H} )/U({\cal H}_+|{\cal H}_+), \quad 
   \Omega_s={i\over 4} \Str\Phi d \Phi\wedge d\Phi.
.\eeq

The reader can now see how this is  related to our 
system, from the experience we have in the previous cases.
In our problem 
we have a free  action which has 
bosons and fermions,
\beq
S_0=\int dx^+dx^-\Big( {1\over 2} \phi^T(-2\partial_-)\partial_+\phi
+{1\over 2}i2\sqrt{2} \psi^T\partial_+\psi-
{1\over 2} m_B^2\phi^T \phi-{1\over 2} \sqrt{2}m_F^2\psi^T{1\over i\partial_-}
\psi\Big)
.\eeq
This action is written in the standard light-cone frame and
one of the components of the Majorana field has been 
eliminated in favor of the other. The transpose refers to
the color indices for the gauge group $SO(N_c)$. 
As it stands this does not require  the full content of the super-geometry,
but as we have seen the 
interaction terms, given by the proper bilinears 
of field operators, makes the use of 
super geometry most convenient:  
when we reformulate our theory in terms of
bilinears, we need the combinations  which can only be expressed in terms 
of odd operators. We will now see   that the Poisson algebra of these 
bilinears can only be formulated as a super-two form.
Moreover  a simple iterative solution of the 
constraint equation reveals that the 
bosonic operators should be given as an 
infinite series of products of odd operators, this is 
why  we think it is most natural to use the full content 
of the Berezin's 
super-analysis.
(We hope to come back to the more mathematical aspects of our 
system  in a future publication).

In  our  theory we have a super-symplectic form and a 
super-quadratic form 
which is in the standard representation given by  
\beq 
  \omega_s=\pmatrix{ -2\partial_-&0\cr 0& i2\sqrt{2}}I_c, \quad
Q_s=\pmatrix{m_B^2&0\cr 0&-\sqrt{2}m_F^2i\partial_-^{-1}}I_c
,\eeq
where we have the identity $I_c$ in the color space.
The relevant operator is
\beq
   \omega_s^{-1}Q_s=\pmatrix{-{m_B^2\over 2}\partial_-^{-1}&0\cr
   0& -{m_F^2\over 2} \partial_-^{-1}}I_c
.\eeq
The complex structure becomes(we drop the identity 
in the color space),
\beq
    J_s=[(-\omega_s^{-1}Q_s)^2]^{-1/2}\omega_s Q_s=\pmatrix{
-(-\partial_-^2)^{1/2}\partial_-^{-1}&0\cr 0&
-(-\partial_-^2)^{1/2}\partial_-^{-1}}
.\eeq
Clearly we can use the Fourier decomposition of our 
fields to diagonalize this and the frequency operator
$K_s$,
\beq
     \phi^\alpha(x^-)=\int {[dp]\over \sqrt{2|p|}} w^\alpha(p)e^{-ipx^-}\quad 
  \psi^\alpha(x^-)=\int {[dp]\over 2^{3/4}}\zeta^\alpha(p)e^{-ipx^-}
,\eeq
here we should think of $w^\alpha$ as even and  $\zeta^\alpha$ 
 odd elements of the Grassmann algebra defined by a 
series (an infinite one) of unspecified generators $\theta^\alpha(p)$,  
Apply the operator $J_s$ to the vectors of this graded 
space,
\beq
     J_s\Psi=J_s\pmatrix{\phi^\alpha\cr \psi^\alpha}=\int [dp] e^{-ipx^-}
(-i\sgn(p))\pmatrix{  (\sqrt{2|p|})^{-1}w^\alpha(p)\cr
   2^{-3/4}\zeta^\alpha(p)}
,\eeq
which we should rewrite as,
\beq 
  (J_s\Psi)(x^-)=\int_0^\infty [dp]e^{-ipx^-}\Big[(-i)
\pmatrix{(\sqrt{2p})^{-1}\bar z^\alpha(p)\cr
2^{-3/4}\bar \xi^\alpha(p)}+i\pmatrix{(\sqrt{2p})^{-1}z^\alpha(p)
\cr 2^{-3/4}\xi^\alpha(p)}\Big]
,\eeq
defining the super-holomorphic coordinates,
$(z^\alpha(p), \xi^\alpha(p))$. 
If we assign now our creation and annihilation operators 
according to the sign of $i$, 
\beq
   \pmatrix{z^\alpha(p)\cr \xi^\alpha(p)}\mapsto \pmatrix{a^{\alpha\dag}(p)\cr 
\chi^{\alpha\dag}(p)}, \quad  \pmatrix{\bar z^\alpha(p)\cr \bar \xi^\alpha(p)}
\mapsto \pmatrix{a^{\alpha}(p)\cr 
\chi^{\alpha}(p)}
,\eeq
we get the  commutation/anticommutation relations for 
$a^{\alpha\dag}(p), a^\beta(q)$ and $\chi^{\alpha\dag}(p),\chi^\alpha(q)$
respectively, and the zero commutator between the two sets.
These commutation/anticommutation relations have   
  the 
operator $\hat \omega_s$ on the right hand-side, this 
is what determines the algebra.
Hence we see that we are in the geometric setting 
we were describing. 
Our bilinears combined in the form of 
$\Phi$ satisfy all the Lie algebra properties.
In fact it is instructive to write down 
the super-symplectic form $\Omega_s$ 
with $\Phi$ expressed in terms of the bilinears $B,F,C,C^\dag$.
Then the reader can see that we have the same 
Poisson brackets satisfied by these 
bilinears.

From the above  discussion  
we again see the remarkable fact that the geometry which is 
defined by the complex structure $J_s$ is independent 
of the parameters of the theory in this 
light-cone method. This means  
{\it even though the masses change due to the interactions this will 
not change the representation of canonical commutation/anticommutation
 relations we started with}, as a result  the geometry stays the same.

Our bilinears will 
correspond to the
generators of the 
automorphisms of this full algebra of 
commutation/anticommutation relations, and it is 
the restricted real $OSp$ group(for finite dimensional
automorphism groups  see
\cite{grosse},  for Bogoliubov automorphisms 
of qausi-free representations see \cite{langmann1, langmann2, langmann3, 
ekstrand}).
We can check that the bilinears we have satisfy the 
Lie algebra conditions and 
the implementability of these automorphisms 
will imply the convergence conditions.
Therefore the evolution of the  system in the 
large-$N_c$ limit  realizes all automorphisms of the 
the quasi-free second quantization of this system
when we think of it without the color part--the color part 
has been averaged out and reduced the system to this 
bilinears. We may give  an  argument 
using the super-coherent states\cite{gradechi1, gradechi2}, similar to the 
ordinary cases: there is a central extension of the 
automorphism group $\widehat{OSp}_1$ which is realized by these 
bilinears on the full Fock space. When we think about the
projective Fock space this descends to the  $OSp_1$ 
group. The orbit of the vacuum 
under this group gives us a classical phase space albeit a
more general one, with a super symplectic form.
The large-$N_c$ limit provides this reduction to 
the space of super-coherent states.
This is  a natural classical phase space and  
the  large-$N_c$ limit corresponds to  this  classical limit.

Before ending our 
discussions we would like to make 
a few comments of general nature.
Let us write down a super-dynamical system in the 
Hamiltonian form
\beq
   S_0= \int dt {1\over 2} \Psi^\tau \omega_s\partial_t  \Psi-
\int dt {1\over 2} \Psi^\tau Q_s \Psi
.\eeq
We assume that the action is an element of the even 
part of the Grassmann algebra.
If we want this to be real we demand 
$(\Psi^\tau Q_s \Psi)^*=\Psi^\tau \tilde E  Q_s^*\Psi=\Psi^\tau Q_s \Psi$,
that is $Q_s=\tilde E  Q_s^*$, we are again using 
$\tilde E=\pmatrix{1&0\cr 0&-1}$ in the standard decomposition.
For the first term it implies the same
$\tilde E \omega_s^*=\omega$.
If we further note that it should be invariant 
under the transpose,
we get for the first term using an integration by parts 
for the time derivative,
$\Psi^\tau \omega_s\partial_t\Psi=-\Psi^\tau \omega_s^\tau \tilde E 
\partial_t \Psi$, which implies $\omega_s=-\omega_s^\tau\tilde E  $
and the second term requires 
$\Psi^\tau Q_s\Psi=\Psi^\tau Q^\tau_s \tilde E \Psi$,
which means we should have $Q_s=Q_s^\tau\tilde E$.
The equations of motion will give us,
\beq
     \partial_t\Psi=\omega_s^{-1} Q_s\Psi
.\eeq 
This suggests that we should further investigate 
operator $\omega_s^{-1}Q_s$ which is a 
type $(1,1)$ tensor, thus a 
true linear transformation. We note that $\omega_s^{-1}Q_s$ is real:
$(\omega_s^{-1}Q_s)^*=(\omega_s^{-1})^*Q_s^*=
\omega_s^{-1}\tilde E \tilde E Q_s=\omega_s^{-1}Q_s$, 
by using the conjugation  properties of $\omega_s$ and $Q$.
This operator  is antisymmetric with respest to the 
form defined by $Q_s$:
\beq   
    Q_s^{-1}(\omega_s^{-1} Q_s)^\tau Q_s=Q_s^{-1}Q_s^\tau(\omega_s^\tau)^{-1}
Q_s=-Q_s^{-1}Q_s\tilde E\tilde E \omega_s^{-1} Q_s= 
-\omega_s^{-1}Q_s,
\eeq
as well as under $\omega_s$.
It would be most natural if we  could use  a generalization of the
polar decomposition for $Q_s^{-1}\omega_s$,
and write this operator as 
$\omega_s^{-1} Q_s=J_s K_s$, where 
$J_s^tJ_s=1$, and  $K_s>0, K_s^t=K_s$, with an 
appropriate transpose $\ ^t$ and positivity 
is assumed to be given a meaning in this super-context. Then we could 
claim that the basis in which $J_s$ is diagonal, 
will tell us the separation of creation and annihilation 
operators in this full generality. 
This can be done in the simple case we looked at, when the operators 
involved only had body parts, and no Grassmann numbers.
Unfortunately for the 
general case we do not have the proper 
mathematical machinery. 
If we could find a super-transformation $S$,
such that $S^{-1} \omega_s^{-1}Q_s S$ is diagonal with each entry 
$(\pm i \lambda_k)$ for a pure number $\lambda_k$ we could postulate 
the quantization by means of canonical commutation/anticommutation 
relations. To the best of our knowledge there is no
such theorem.
We think these questions deserve further investigations.

\section{Acknowledgements}

This research is largely based on the ideas of S. G. Rajeev, we would like to 
express our gratitute to him for his generosity and being such a 
unique source  of inspiration and guidance.
O. T. Turgut's research in Stockholm has been supported by 
G. Gustafsson fellowship. O. T. Turgut would like to   
thank especially to J. Mickelsson for such a great 
opportunity to be in KTH, and he thanks to J. Mickelsson, E. Langmann
for hospitality of the group and for discussions.
E. Langmann made a critical reading of the 
paper and improved the presentation drastically.
We thank   Bogazici University 
Physics Department for the support and the excellent  
atmosphere. 
We also would  like to acknowledge discussions with 
M. Arik, J. Gracia-Bondia, A. Konechny, Y. Nutku and C. Saclioglu,
W. Zakrzewski.

\end{document}